\documentclass[11pt]{article}
\usepackage{tikz}
\usepackage[colorinlistoftodos,textsize=footnotesize]{todonotes}
\usetikzlibrary{matrix,arrows,shapes}
\usepackage{pgf}
\usepackage{cancel}
\usepackage{url}
\usepackage{longtable}
\usepackage{enumerate}
\usepackage{amsthm}
\usepackage{amssymb}
\usepackage{amsmath}
\usepackage{amscd}
\usepackage{eucal}
\usepackage{mathrsfs}
\usepackage{upgreek}
\usepackage{mathtools}
\usepackage{dsfont}
\usepackage{yfonts}
\usepackage[T1]{fontenc}
\usepackage{bbm}
\usepackage{geometry}
\usepackage{graphbox}
\usepackage{array}
\usepackage[retainorgcmds]{IEEEtrantools}
\usepackage{booktabs}
\usepackage{enumerate}
\usepackage{feynmf}

\usetikzlibrary{trees}
\usetikzlibrary{decorations.pathmorphing}
\usetikzlibrary{decorations.markings}
\tikzset{
	photon/.style={decorate, decoration={snake}, draw=red},
	electron/.style={draw=blue, postaction={decorate},
		decoration={markings,mark=at position .55 with {\arrow[draw=blue]{>}}}},
	gluon/.style={decorate, draw=magenta,
		decoration={coil,amplitude=4pt, segment length=5pt}},
	sderiv/.style={postaction={decorate},
		decoration={markings,mark=at position .3 with {\arrow{>}}}},
	tderiv/.style={postaction={decorate},
		decoration={markings,mark=at position .7 with {\arrow{<}}}},
	stderiv/.style={postaction={decorate},
		decoration={markings,mark=at position .7 with {\arrow{<}},mark=at position .3 with {\arrow{>}}}}
}
\usepackage{scalerel}
\usepackage{stackengine}

\definecolor{see}{RGB}{67,75,179}
\definecolor{darksee}{RGB}{42,44,148}
\definecolor{honey}{RGB}{232,180,129}
\definecolor{lighthoney}{RGB}{255,254,220}
\definecolor{citecol}{rgb}{0.5,0,0} 
\definecolor{blue1}{RGB}{130,150,209}
\usepackage{graphicx,stackengine,scalerel}

\definecolor{blue1}{RGB}{130,150,209}
\definecolor{blue2}{RGB}{42,60,122}
\usepackage{soul}
\usepackage[colorinlistoftodos,textsize=footnotesize]{todonotes}

\setstcolor{red}
\definecolor{see}{RGB}{67,75,179}
\usepackage{hyperref}
\hypersetup
{
	bookmarks=true,
	bookmarksopen=true,
	pdfmenubar=true, %menubar shown
	pdfhighlight=/O, %effect of clicking on a link
	colorlinks=true, %couleurs sur les liens hypertextes
	pdfpagemode=None, %aucun mode de page
	pdfpagelayout=SinglePage, %ouverture en simple page
	pdffitwindow=true, %pages ouvertes entierement dans toute la fenetre
	linkcolor=see, %couleur des liens hypertextes internes
	citecolor=red!75!darkgray, %couleur des liens pour les citations
	urlcolor=see %couleur des liens pour les url
}
\newlength{\bibitemsep}
\newlength{\bibparskip}
\let\oldthebibliography\thebibliography
\renewcommand\thebibliography[1]{%
	\oldthebibliography{#1}%
	\setlength{\parskip}{\bibitemsep}%
	\setlength{\itemsep}{\bibparskip}%
}
\newcommand{\no}[1]{\mathop{\mathopen: {#1} \mathclose:}}
\def\Caus{\mathbf{Caus}}

\newcommand{\fP}{\mathfrak{P}}

\newcommand{\inj}{{\rm inj}}

\def\Alg{\mathbf{Alg}}
\def\CAlg{\mathbf{CAlg}}
\def\PAlg{\mathbf{PAlg}}
\def\Ch{\mathbf{Ch}}
\def\Caus{\mathbf{Caus}}

\def\Nuc{\mathbf{Nuc}}
\def\TVec{\mathbf{TVec}}
\def\Nuch{\mathbf{Nuc}_{\hbar}}

\def\Cat{\mathbf{C}}

\newcommand{\frakgo}{\mathfrak{g}}

\newcommand{\CE}{\mathcal{CE}}  
\newcommand{\E}{\mathcal{E}}
\newcommand{\V}{\mathcal{V}}

\newcommand{\fA}{\mathfrak{A}}

\newcommand{\F}{\mathcal{F}}

\newcommand{\fG}{\mathfrak{G}}

\newcommand{\BV}{\mathcal{BV}}

\newcommand{\frakg}{\mathfrak{g}}

\newcommand{\euL}{\mathscr{L}}

\newcommand{\Gcal}{\mathcal{G}}  % gauge group
\newcommand{\BVcal}{\mathcal {BV}}

\newcommand{\CEcal}{\mathcal {CE}}

\newcommand{\Dcal}{\mathcal{D}}
\newcommand{\Ecal}{\mathcal{E}} 

\newcommand{\Fcal}{\mathcal{F}}

\newcommand{\Ncal}{\mathcal{N}}
\newcommand{\Mcal}{\mathcal{M}}
\newcommand{\Ocal}{\mathcal{O}}
\newcommand{\Scal}{\mathcal{S}}
\newcommand{\Pcal}{\mathcal{P}}

\newcommand{\Tcal}{\mathcal{T}}
\newcommand{\Vcal}{\mathcal{V}}

\newcommand{\Ci}{\mathcal{C}^\infty} % smooth functions
\newcommand{\WF}{\mathrm{WF}}         % wave front set
\newcommand{\dvol}{\mathrm{dvol}_{\sst{M}}} % volume form
\DeclareMathOperator{\tr}{\mathrm{tr}}                 % trace

\newcommand{\loc}{\mathrm{loc}}
\newcommand{\inv}{\mathrm{inv}}
\newcommand{\reg}{\mathrm{reg}}

\newcommand{\pg}{\mathrm{pg}}

\newcommand{\af}{\mathrm{af}}
\newcommand{\ta}{\mathrm{ta}}
\newcommand{\gh}{\mathrm{gh}}

\newcommand{\mc}{{\mu\mathrm{c}}}
\newcommand{\ml}{\mathrm{ml}}
\newcommand{\ex}{\mathrm{ext}}

\newcommand{\YM}{\rm\sst YM}
\newcommand{\NN}{\mathbb{N}}          % natural naumbers
\newcommand{\RR}{\mathbb{R}}           % real  numbers
\newcommand{\CC}{\mathbb{C}}           % complex numbers

\newcommand{\al}{\alpha}

\newcommand{\De}{\Delta}

\newcommand{\La}{\Lambda}

\newcommand{\ph}{\varphi}

\newcommand{\T}{\cdot_{{}^\Tcal}}

\newcommand{\TT}{\Tcal}

\newcommand{\Poi}[2]{\left\lfloor#1,#2\right\rfloor}
\newcommand{\qme}{{\textsc{qme}}}

\newcommand{\cme}{{\textsc{cme}}}

\newcommand{\sst}[1]{\scriptscriptstyle{#1}}  % small font for the subscripts
\newcommand{\minus}{\sst{-1}}   % power ^{-1}
\newcommand{\1}{\mathds{1}}                         % identity
\newcommand{\pa}{\partial}                              % partial derivative
\newcommand{\be}{\begin{equation}}
\newcommand{\ee}{\end{equation}}

\newcommand{\Lap}{\bigtriangleup}

\DeclareMathOperator{\supp}{supp}      % support

\newcommand{\Pei}[2]{\lfloor #1, #2 \rfloor}
\newcommand{\skal}[2]{\left< #1 , #2 \right>}

\newcommand{\dgr}{{\sst\ddagger}}

\newcommand{\cinfty}{\mathcal{C}^{\infty}}
\theoremstyle{plain}
\newtheorem{thm}{Theorem}[section]

\theoremstyle{definition}
\newtheorem{df}[thm]{Definition}
\theoremstyle{remark}

\newtheorem{exa}[thm]{Example}

	\title{Perturbative algebraic quantum field theory and beyond}
	\date{}
\author{Romeo Brunetti, Klaus Fredenhagen, Kasia Rejzner}
	
\begin{document}	
	\maketitle	

\begin{abstract}
    In this review, we summarize the main ideas of perturbative algebraic quantum field theory, which is a rigorous framework combining some of the Haag-Kastler axioms with perturbative methods involving formal power series. It allows for the construction of interacting QFT models in four spacetime dimensions and works on arbitrary globally hyperbolic manifolds. This approach has also led to the development of a non-perturbative construction of local nets of C*-algebras for interacting theories, which will also be discussed at the end of this review.
\end{abstract}

%	\tableofcontents
\section{Introduction}

Algebraic Quantum Field Theory (AQFT), as reviewed in \cite{BuchholzFredenhagenEnzyclopedia}, presents a framework for qualitatively describing a broad spectrum of phenomena within particle physics and certain domains of solid-state physics. However, there is a prevailing notion that AQFT's formalism must be relinquished to establish substantial connections with experimental observations. To date, constructing a single model of interacting AQFT in 4-dimensional Minkowski space remains elusive, an unfortunate fact shared with other rigorous approaches to Quantum Field Theory. 
Free theories, i.e. field theories with linear equations of motion, can be constructed, and one tries to treat interacting theories as deformations of free theories. But the interactions which are compatible with relativistic causality are too singular for a treatment by the presently available operator algebraic methods. 

Conventional Quantum Field Theory (QFT) textbooks typically approach this problem by adopting one of two primary methodologies. The first involves commencing with canonical quantization of free field theory on Fock space and endeavoring to construct the interacting theory within the interaction picture. Alternatively, they may employ the path integral formalism by deforming a Gaussian measure corresponding to the free theory. In both cases, 
one obtains a description of the theory in terms of formal power series, and in addition to the quite nontrivial problem to compute the terms of the series one is left with the problem to truncate the series in a meaningful way in order to
compare the theory with experiments.

% a fundamental role is played by the \emph{time-ordered products}.

Both approaches, albeit heuristic, have been refined by efforts of successive generations of physicists, yielding computationally tractable formalisms. Despite certain challenges, such as infrared problems and divergences, these formalisms have displayed remarkable success in yielding experimental agreement.

The path integral approach, reminiscent of probability theory after employing imaginary time (Wick rotation), offers computational advantages particularly evident in momentum space integrals corresponding to Feynman diagrams, but causality is less tractable, as it arises from non-commutativity of the operator product, which is less evident at the level of Wick-rotated vacuum expectation values. Conversely, the canonical approach directly involves operator products but struggles with a consistent definition of time-ordered products. Moreover, in both formalisms, the extension to curved spacetimes is problematic.

Causal perturbation theory \cite{EG}, inspired by earlier works \cite{StuRiv,BS}, offers a \emph{complete} solution to these problems by defining time-ordered products as operator-valued distributions. These are defined up to coinciding points and the usual ambiguities at those points are completely characterized. For more details and references see \cite{MichaelChapter}. In this framework one observes that the algebraic structure of the interacting theory can, to a large extent, be determined, while the expectation values remain formal power series. This suggests to consider the arising framework as a modified version of the Haag-Kastler axioms. 

This connection to the Haag-Kastler axioms can be obtained by either weakening the assumption that local algebras are $C^*$-algebras \cite{BF0} and allowing for both algebras and states to be formal power series in $\hbar$, or by using relations arising in causal perturbation theory as a motivation for building a certain net of $C^*$-algebras \cite{BF19}, with the caveat that the existence of physically distinguished states (e.g. the vacuum state) is not yet known. Both approaches work also on curved spacetimes.

\section{Weakening the AQFT axioms}

In this section, we investigate the potential of relaxing certain assumptions inherent in the Haag-Kastler framework, thereby accommodating models that to date are known only in the perturbative setting. This modified framework is termed \textit{perturbative Algebraic Quantum Field Theory} (pAQFT). The extension of the Haag-Kastler axioms to the perturbative realm has been elaborated upon in \cite{BF0,DF,DF02,DF04,DF05,DFloop,BreDue,Boas,BoasDue,BDF,Rej11a}. 

The extension of the Haag-Kastler framework to curved spacetime has historically evolved as a separate line of inquiry. Significant early contributions include seminal works by \cite{Kay78,Dim,KW91,Dim92}. Subsequently, these independent developments converged with the emergence of pAQFT on curved spacetimes, catalyzed by a series of publications \cite{BFK96,BF97,BF0,BFV,HW,HW01,HW02,HW05}. The crucial step was to understand renormalization on curved spacetime  using the principle of general local covariance, which is the fundamental principle in AQFT on curved spacetimes \cite{BFV}.  More on QFT on curved spacetime can be found in \cite{BernardChapter}. For further reading on the subject of pAQFT, we recommend the following review books: \cite{Book,Due19}.

Abelian gauge theories were subsequently addressed in \cite{DFqed}, while Yang-Mills theories are discussed in a later work by Hollands \cite{H}. 
Simultaneously, the theoretical underpinnings of perturbative Algebraic Quantum Field Theory (pAQFT) saw advancements, particularly through the functional approach \cite{DF02,DF04,BDF}, a methodology also adopted in this text. 

The incorporation of the Batalin-Vilkovisky (BV) formalism into the pAQFT framework was explored in papers by Fredenhagen and Rejzner \cite{FR,FR3,Rej11b}. This extension broadens the scope of pAQFT to encompass theories featuring local symmetries, such as the Yang-Mills theories, the bosonic string \cite{BRZ}, and effective quantum gravity \cite{BFRej13}.

An elegant formulation of the pAQFT axioms is provided using the language of category theory (see also \cite{BFV} for the category-theoretic formulation of AQFT on curved spacetimes), so we will adopt it in this review as well. We mostly follow the conventions and notations of \cite{GR20}. For readers unfamiliar with category theory, it is enough to think about a category as a collection of objects and maps between them, called morphisms. A functor between two categories constitutes a map on objects and a compatible map on morphisms, which maps identity morphisms to identity morphisms and preserves the composition.

Here are some important categories that we will use to phrase the pAQFT axioms. 
Let $\Nuc$ represent the category comprising nuclear, topological locally convex vector spaces, a subset of the broader category of topological locally convex spaces $\TVec$. Being nuclear means that there is no ambiguity in defining a tensor product, so $\Nuc$ comes equipped with a natural monoidal structure $\widehat{\otimes}$.

The category of unital associative algebras within such vector spaces will be denoted as $\Alg(\Nuc)$ and its morphisms consist of continuous linear algebra morphisms. Correspondingly,  $\CAlg(\Nuc)$ is used to denote unital commutative algebras within $\Nuc$, while $\PAlg(\Nuc)$ represents unital Poisson algebras therein. Typically, we require also an involution compatible with multiplication, for which we employ $\Alg^*(\Nuc)$, $\CAlg^*(\Nuc)$, and $\PAlg^*(\Nuc)$, respectively.

 Given an additive category $\Cat$, we employ $\text{Ch}(\Cat)$ to represent the category of cochain complexes and cochain maps within $\Cat$. We use the superscript ``$\inj$,'' if we want to impose the injectivity condition on morphisms. This will be important for the formulation of the \textit{Isotony axiom} in the Haag-Kastler framework. The subscript $\hbar$ signifies that we work with formal power series. For a precise definition, see section 2.1.3 of \cite{GR20}.

In the notations above, $\Alg^*(\Nuc_\hbar)^\inj$ is the category whose objects are formal power series with coefficients in nuclear, topological locally convex unital $*$-algebras and whose morphisms are formal power series in \emph{injective} continuous algebra morphisms. 

Let $\Mcal=(M,g)$ be a globally hyperbolic (meaning that it has a Cauchy surface), oriented and time-oriented spacetime and let $\Caus(\Mcal)$ be the collection of relatively compact, connected, contractible, causally convex subsets $\Ocal\subset \Mcal$. Note that the inclusion relation $\subset$ is a partial order on $\Caus(\Mcal)$ making $(\Caus(\Mcal),\subset~)$ into a poset and hence a category in its own right.

It is convenient to put classical field theory into the same framework as QFT. We will use the following definition.
\begin{df}\label{ClassFT}
	A \emph{classical field theory model} on a spacetime $\Mcal$ is a functor $\fP : \Caus(\Mcal) \to \PAlg^*(\Nuc)^\inj$ that obeys \emph{Einstein causality}, i.e.:
	for $\Ocal_1,\Ocal_2\in \Caus(\Mcal)$ that are spacelike to each other, we have
	\[
	\Poi{\mathfrak{P}(\Ocal_1)}{\mathfrak{P}(\Ocal_2)}_{\Ocal}=\{0\}\,,
	\]
	where $\Poi{.}{.}_{\Ocal}$ is the Poisson bracket in  any $\fP(\Ocal)$ for an $\Ocal$ that contains both $\Ocal_1$ and $\Ocal_2$.
\end{df}
Dynamics is encoded in the notion of being \emph{on-shell}. In practice, for classical field theory, it means that one works on the space of solutions to equations of motion. More abstractly, we invoke the following definition.
\begin{df}
	A model is said to be \emph{on-shell} if in addition it satisfies the \emph{time-slice axiom}:
	for any $\Ncal\in\Caus(\Mcal)$ a neighborhood of a Cauchy surface in the region $\Ocal\in\Caus(\Mcal)$, 
	then $\mathfrak{P}$ sends the inclusion $\Ncal \subset \Ocal$ to an isomorphism $\mathfrak{P}(\Ncal)\cong \mathfrak{P}(\Ocal)$. Otherwise the model is called \emph{off-shell}.
\end{df}

Next, we define the quantum theory, together with the appropriate on-shell version.
\begin{df}\label{AQFT}
	A \emph{pAQFT model} on a spacetime $\Mcal$ is a functor $\mathfrak{A} : \Caus(\Mcal) \to \Alg^*(\Nuc_\hbar)^{\inj}$ that satisfies \emph{Einstein causality} (Spacelike-separated observables commute). That is, for $\Ocal_1,\Ocal_2\in \Caus(\Mcal)$ that are spacelike to each other, we have
	\[
	[\fA(\Ocal_1),\fA(\Ocal_2)]_\Ocal=\{0\}\,,
	\]
	where $[.,.]_{\Ocal}$ is the commutator in  any $\fA(\Ocal)$ for an $\Ocal$ that contains both $\Ocal_1$ and $\Ocal_2$. 
\end{df}

\begin{df}\label{timeslice}
	A pAQFT model is said to be \emph{on-shell} if in addition it satisfies the \emph{time-slice axiom} (where one simply replaces $\mathfrak{P}$ by $\fA$ in the definition above).
	Otherwise, it is \textbf{off-shell}. 
\end{df}

The definitions above are appropriate for describing theories without local symmetries, e.g. the scalar field, as was done in \cite{BDF}. In order to encompass symmetries and allow models including Yang-Mills theory and effective gravity, one needs to weaken the axioms even further. A convenient description uses homological algebra. In the simplest scenario, one replaces algebras with chain complexes constructed from \emph{differential graded algebras} (dga). A dga is an algebra equipped with grading and a differential, i.e. a map that squares to zero, satisfies the graded Leibniz rule and changes the grade by 1 (co-chain complexes) or -1 (chain complexes). We will use the abbreviation ``dg'' to denote \emph{differential graded}. 

\begin{df}\label{dgClassFT}
	A \emph{semistrict dg classical field theory model} on a spacetime $\Mcal$ is a functor $\mathfrak{P} : \Caus(\Mcal) \to \PAlg^*(\Ch(\Nuc))$,  so that each $\mathfrak{P}(\Ocal)$ is a locally convex dg Poisson $*$-algebra satisfying \emph{Einstein causality}: for $\Ocal_1,\Ocal_2\in \Caus(\Mcal)$ that are spacelike to each other, the bracket $\Poi{\mathfrak{P}(\Ocal_1)}{\mathfrak{P}(\Ocal_2)}$ vanishes at the level of cohomology
	in $\fP(\Ocal')$ for any $\Ocal' \in \Caus(\Mcal)$ that contains both $\Ocal_1$ and~$\Ocal_2$.
	
	It satisfies the \emph{time-slice axiom} if for any $\Ncal\in\Caus(\Mcal)$ a neighborhood of a Cauchy surface in the region $\Ocal\in\Caus(\Mcal)$, 
	then the map $\mathfrak{P}(\Ncal) \to \mathfrak{P}(\Ocal)$ is a quasi-isomorphism.
\end{df}

More explicitly, given $A\in \mathfrak{P}(\Ocal_1)$ and $B\in \mathfrak{P}(\Ocal_2)$ for spacalike $\Ocal_1,\Ocal_2\in \Caus(\Mcal)$, we have that $\Poi{A}{B}$ is in the image of $d$, where $d$ is the differential operator of the differential graded algebra $\fP(\Ocal')$, where $\Ocal'$ contains both $\Ocal_1$ and $\Ocal_2$.

\begin{df}\label{LCQFT}
	A \emph{semistrict dg QFT model} on a spacetime $\Mcal$ is a functor 
	$\fA : \Caus(\Mcal) \to \Alg^*(\Ch(\Nuch))$,  so that each $\fA(\Ocal)$ is a locally convex unital $*$-dg algebra satisfying \emph{Einstein causality}: spacelike-separated observables commute at the level of cohomology. 
	That is, for $\Ocal_1,\Ocal_2\in \Caus(\Mcal)$ that are spacelike to each other, the bracket
	$[\fA(\Ocal_1),\fA(\Ocal_2)]$ is exact in $\fA(\Ocal')$ for any $\Ocal' \in \Caus(\Mcal)$ that contains both $\Ocal_1$ and~$\Ocal_2$.
	
	It satisfies the \emph{time-slice axiom} if for any $\Ncal\in\Caus(\Mcal)$ a neighborhood of a Cauchy surface in the region $\Ocal\in\Caus(\Mcal)$, 
	then the map $\fA(\Ncal) \to \fA(\Ocal)$ is a quasi-isomorphism.
\end{df}

An even more general setting is provided by the \emph{homotopy AQFT}, which is covered in \cite{AlexAndMarcoChapter}.

\section{Constructing models in pAQFT}

\subsection{Kinematical structure}
 The ideas presented here apply to general theories featuring local gauge invariance. However, for illustrative purposes, we focus on the self-interacting Yang-Mills theory.
 
 We denote by $\mathcal{E}$ the configuration space of the theory, construed as the space of smooth sections of a vector bundle $E\xrightarrow{\pi} M$ over $M$. $\mathcal{E}$ defines the type of object the theory encompasses (e.g., scalar fields, tensor fields). To facilitate our discussion, we introduce some notation:
\begin{itemize}
		  \setlength\itemsep{-0.2em}
	\item $\Ecal_c$ -- the space of smooth compactly supported sections of $E$.
	\item $\Dcal$ -- the space of smooth compactly supported functions on $\Mcal$.
	\item  $\Ecal'$,  $\Ecal'_c$ -- complexifications of topological duals of  $\Ecal$ and  $\Ecal_c$ respectively (both with the strong topology).
	\item $\Ecal^*$ -- the space of smooth sections of the  dual bundle $E^*$. 
	\item $\Ecal^!$ -- the complexification of the space of sections of $E^*$ tensored with the bundle of densities over $M$.
	\item $\Ecal^!(M^n)$ -- the complexified space of sections of the $n$-fold exterior tensor product of $E^*$ bundle  tensored with the bundle of densities.
\end{itemize}
Elements of $\Ecal$ are always denoted by $\ph$, even if they carry indices. 

We focus on two running examples:
	\begin{itemize}
			  \setlength\itemsep{-0.2em}
	\item \textit{Scalar field}, with the configuration space $\Ecal=\Ci(M,\RR)$. 
	 \item \textit{Yang-Mills theories}. 
	 We consider $G$, a semisimple compact Lie group, and $\mathfrak{k}$ its Lie algebra. For simplicity, let's opt for the trivial bundle $P=M\times G$ over $M$, and define the off-shell configuration space of the Yang-Mills theory as $\mathcal{E}=\Omega^1(M,\mathfrak{k})$. Locally, one can always work with trivial bundles and the problem of gluing local theories together can be addressed separately, e.g. using a Kan extension \cite{BSW19} or the language of factorisation algebras \cite{GR22}.
	\end{itemize}

We characterize classical observables as functionals defined on the configuration space $\mathcal{E}$. To establish a rigorous mathematical foundation, we endow $\mathcal{E}$ with its natural Fr\'echet topology and consider the space of (Bastiani) smooth functionals $\mathcal{C}^\infty(\mathcal{E},\mathbb{C})$, as defined in \cite{Bas64}. These functionals constitute our observables.

From a physical standpoint, an observable in classical theory assigns a numerical value to a given field configuration, corresponding to a measurement outcome (e.g., energy density at a specific spacetime point). The prerequisite of smoothness ensures the well-definedness of all algebraic structures we aim to introduce on these observables.

The particular case of polynomial functionals of the scalar field on Minkowski spacetime is discussed in \cite{MichaelChapter}.

An essential concept to consider next is the notion of \textit{spacetime support} of a functional, which encapsulates the localization characteristics of observables. Additionally, we emphasize the significance of \textit{additivity} as another fundamental property, which encodes locality.
\begin{df}
	The spacetime support of a functional is defined by
	\begin{align}\label{support}
	\supp\, F=\{ & x\in M|\forall \text{ neighbourhoods }U\text{ of }x\ \exists \ph_1,\ph_2\in\E, \supp\, \ph_2\subset U 
	\\ & \text{ such that }F(\ph_1+\ph_2)\not= F(\ph_1)\}\ .\nonumber
	\end{align}
\end{df}
\begin{df}
	A functional $F$ is called \textit{additive} if
	\be\label{L:add}
	F(\ph+\chi+\psi)=F(\ph+\chi)-F(\chi)+F(\chi+\psi)\,,
	\ee
	for $\ph+\chi+\psi\in\Ecal$ and $\supp\,\ph\cap\supp\,\psi=\emptyset$. 
\end{df}
In the physics literature, typically a functional is called \textit{local}  if it can be expressed as:
\[
F(\ph)=\int\limits_M \omega(j^k_x(\ph))\,d\mu(x)\,,
\]
where $\omega$ is a function on the jet bundle over $M$ and $j^k_x(\ph)=(x,\ph(x),\pa \ph(x),\dots)$, with derivatives up to order $k$, is the $k$-th jet of $\ph$ at the point $x$.

In \cite{BDGR}, building upon concepts introduced in \cite{BFR}, it was demonstrated that local functionals can be identified as smooth functionals satisfying \eqref{L:add} and possessing smooth first derivatives. Further features of additivity and its extensions were explored in \cite{Rej19}.

We denote the space of compactly supported smooth local functions on $\mathcal{E}$ as $\mathcal{F}_\text{loc}$. By completing $\mathcal{F}_\text{loc}$ with respect to the pointwise product given by $F\cdot G(\phi)=F(\phi)G(\phi)$, we obtain the commutative algebra $\mathcal{F}$ of \textit{multilocal functionals}.

Additionally, we introduce the concept of \textit{regular functionals}. A functional $F$ belongs to $\mathcal{F}_{\text{reg}}$ if all its derivatives $F^{(n)}(\phi)$ are smooth, meaning that for every $\phi \in \mathcal{E}$ and $n \in \mathbb{N}$, we have $F^{(n)}(\phi) \in \mathcal{E}^!(M^n)$.

For simplicity, in this work all the functionals that we consider are also assumed to be polynomial. In general, one needs to use some additional assumption on functional derivatives e.g. equicontinuity in $\ph$, otherwise the relevant algebraic operations might fail to remain within a given class of functionals. This issue has been recently identified and addressed in \cite{HRV}.
\subsection{Dynamics and symmetries}

\subsubsection{Dynamics}

To introduce dynamics, we employ a generalized version of the Lagrangian formalism, as outlined in \cite{BDF}. Ideally, we aim to derive the equations of motion and symmetries from the action principle. However, a potential challenge arises due to the non-compact nature of the manifolds we work with. For instance, the integral of a Lagrangian density such as $\frac{1}{2}(\nabla^\mu\phi \nabla_\mu\phi -m^2 \phi^2)$ over the entire manifold $M$ does not converge if $\phi$ lacks compact support. One might consider restricting attention to compactly supported configurations, but this approach fails since the desired equations of motion typically lack non-trivial compactly supported solutions. To circumvent this issue, we multiply the Lagrangian density with a cutoff function $f\in\Dcal\equiv\Ci_c(M,\RR)$ and define all relevant quantities (e.g., the Euler-Lagrange derivative) in a manner independent of $f$. To formalize this approach, we introduce the concept of a \textit{generalized Lagrangian}.
\begin{df}\label{Lagr}
	A \textit{generalized Lagrangian} on a fixed spacetime $\Mcal$ is a map $L:\Dcal\rightarrow\Fcal_{\loc}$ such that
	\begin{enumerate}[i)]
		%\item $L(f)$ is selfadjoint (i.e. real-valued) for all $f\in\euD$,
		\item $L(f+g+h)=L(f+g)-L(g)+L(g+h)$ for $f,g,h\in\Dcal$ with $\supp\,f\cap\supp\,h=\varnothing$ ({\bf Additivity}).
		\item $\supp(L(f))\subseteq \supp(f)$ ({\bf Support}).
		\item Let $\Gcal$ be the isometry group of the spacetime $M$ (for Minkowski spacetime we set $\Gcal$ to be the proper orthochronous Poincar\'e group $\Pcal^\uparrow_+$.). We require that $L(f)(g^*\ph)=L(g_*f)(\ph)$ for every $g\in\Gcal$ ({\bf Covariance}).
	\end{enumerate}
Let $\euL$ denote the space of all generalized Lagrangians 
\end{df}
Now we identify generalized Lagrangians that differ by a total derivative.
\begin{df}[\cite{BDF}] 
	Actions $S(L)$ are defined as  equivalence classes of Lagrangians, where two Lagrangians $L_1,L_2$ are called equivalent $L_1\sim L_2$  if
	\be\label{equ}
	\supp (L_{1}-L_{2})(f)\subset\supp\, df\,, 
	\ee
	for all $f\in\Dcal$. 
\end{df}
\begin{exa}
	The generalized Lagrangian of the free scalar field is
\[
L_0(f)[\ph]=\frac{1}{2}\int_M (\nabla^\mu\ph \nabla_\mu\ph -m^2 \ph^2)f d\mu_g\,.
\]
For the Yang-Mills theory, we have
\[
L_{\YM}(f)[A]=-\frac{1}{2}\int_M f\,\tr(F \wedge * F)\,,
\]
where $F=dA+\frac{i\lambda}{2}[A,A]$, $A\in\Ecal$,  $\lambda$ is the coupling constant, $*$ is the Hodge operator and $\tr$ is the trace in the adjoint representation, given by the Killing-Cartan metric $\kappa$.
\end{exa} 
%More explicitly, denoting the generators of $g$ by $T_I$, $I\in \{1,\dots,N\}$ ($N$ being the dimension of the gauge group), we can write $A=A^I_\mu T^I dx^\mu$ and express $F=F^I_{\mu\nu} T_I dx^\mu\wedge dx^\nu$, using the structure constants $f_{IJ}^K$:
%\[
%F^I_{\mu\nu}=\nabla_\mu A^I_\nu-\nabla_\nu A^I_\mu+i\lambda f_{JK}^IA_\mu^JA_\nu^K\,.
%\]
%Here we use the convention $[T_I,T_J]=f_{IJ}^K T_K$.

Following \cite{BF19}, we define the finite variation of a generalized Lagrangian. 
\begin{df}\label{df:delta:L}
	Let $L\in \euL$, $\ph\in\Ecal$. Define a functional $\delta L:\Dcal\times\Ecal\rightarrow \RR$ by
	\[
	\delta L(\psi)[\ph]\doteq L(f)[\ph+\psi]-L(f)[\ph]\,, 
	\]
	where $\ph\in\Ecal$, $\psi\in \Ecal_c$ and $f\equiv 1$ on $\supp \psi$ (the map $\delta L(\psi)[\ph]$ thus defined does not depend on the particular choice of $f$).
\end{df}
The infinitesimal version of this variation results in the \textit{Euler-Lagrange derivative} of $S$. The equations of motion are interpreted following the framework established in \cite{BDF}. Specifically, the Euler-Lagrange derivative of $S$ is a 1-form on $\mathcal{E}$, denoted by $dS: \mathcal{E} \to \mathcal{E}_c'$, defined by
\be\label{ELd}
\left<dS(\ph),\psi\right>\doteq \lim_{t\rightarrow 0}
\tfrac{1}{t}\delta L(t\psi)[\ph]=\int\frac{\delta L(f)}{\delta \ph (x)} \psi(x) \,,
\ee
with $\psi\in\Ecal_c$ and $f\equiv 1$ on $\supp \psi$, and $\frac{\delta L(f)}{\delta \ph}$ is understood as an element of $\Ecal^!\subset \Ecal_c'$. 
The field equation is expressed by the condition on $\phi$ as follows:
\begin{equation}
	dS(\phi) \equiv 0 \ .\label{eom}
\end{equation}
Therefore, from a geometric perspective, the solution space corresponds to the \textit{zero locus of the 1-form $dS$}. Let $\mathcal{E}_S \subset \mathcal{E}$ denote the space of solutions to \eqref{eom}. Our focus lies on the space $\mathcal{F}_S$ of functionals defined on $\mathcal{E}_S$, which we refer to as \textit{on-shell} functionals.
\begin{exa} Examples of equations of motion:	
	\begin{itemize}
			  \setlength\itemsep{-0.2em}
\item	\textbf{Free scalar field}:
	$dS_0(\ph)=-(\Box+m^2)\ph$,
	where $\Box$ is the wave operator (d'Alembertian).
\item \textbf{Yang-Mills theory}:	
$dS_{\YM}(A)=D_A\!*\!F$,
where $D_A$ is the covariant derivative induced by the connection $A$. 
\end{itemize}
\end{exa}
For systems with several fields (or components), we use the notation $\frac{\delta S}{\delta \ph^{\alpha}}$ for  $\frac{\delta L(f)}{\delta \ph^{\alpha}}$ evaluated at $f\equiv 1$ and treated as a component of the form $dS$. 

\subsubsection{Symmetries}
Symmetries are defined as directions within the configuration space $\mathcal{E}$, originating from a specific point, along which the action is constant. Geometrically, these are vector fields $X$ on $\Ecal$ such that
\[
\partial_X S\equiv 0\,,
\]
where
\[
\partial_X S\doteq \int\frac{\delta L(f)}{\delta \ph(x)} X(x) \,,\quad f\equiv 1\ \textrm{on}\ \supp X\,,
\]
and $X\in \Gamma(T\Ecal)$ is identified with a map from $\Ecal$ to $\Ecal_c^{\sst \CC}$.

Often we use the formal notation
\[
X=\int  X(x) \frac{\delta}{\delta \ph(x)} \,,
\]
and in order to make contact with the physics literature, we identify the basis on the fiber $T_\ph\Ecal$ as the antifields $\frac{\delta}{\delta \ph(x)}\equiv \ph^\ddagger(x)$.

Let's assume that there exists a Lie-algebra morphism $\rho:\mathfrak{g}_c\rightarrow \Gamma (T\mathcal{E})$. This morphism stems from a specified local action $\sigma$ of some Lie algebra $\mathfrak{g}_c$ on $\mathcal{E}$, characterized by
\[
\rho(\xi)F[\phi] \doteq \left< F^{(1)}(\phi),\sigma(\xi)\phi \right> \equiv \int_M \frac{\delta F}{\delta \phi(x)} \sigma(\xi) \phi(x)\,.
\]
We assume $\mathfrak{g}_c$ to be a space of smooth compactly supported sections of some vector bundle over $M$, with the action $\sigma$ on $\mathcal{E}$ being local.

Assume that all the space of all symmetries is generated in appropriate sense by $\rho(\mathfrak{g}_c)$ and $I$, the ideal generated by symmetries that vanish on the space of solutions (vanish on-shell).

\begin{exa}
For Yang-Mills theory, we have $\frakg_c=\Gamma_c(M,\mathfrak{k})$ and the local action $\sigma$ is 
\[
\sigma(\xi)A\doteq d\xi+[A,\xi]=D_A\xi\,,\quad \xi\in\frakg_c\,.
\]
\end{exa}
The existence of local symmetries implies redundancies in the equations of motion, so the zero locus of $dS$, comprises \textit{orbits} of the action $\sigma$ of $\mathfrak{g}$ on $\mathcal{E}$. 

%A local and compactly supported vector field $X$ can be represented in terms of some differential operator:
%\[
%X^\alpha(x)[\phi] = Q^\alpha_{\ \beta}(\phi) \phi^\beta(x) = a(x)[\phi] \phi(x) + b^\mu(x)[\phi] \nabla_\mu\phi(x) + \dots\,.
%\]
%Here $\alpha$ runs through all the fields of the theory. For self-interacting Yang-Mills, these are just the components of the vector potential, but later on we will see that further fields will be added.

%Thus, the condition for $X$ to be a symmetry can be expressed as:
%\begin{equation}\label{noether}
%	0 = \int \frac{\delta S}{\delta \phi^\alpha(x)}X^\alpha(x) d\mu(x) = \int \phi^\beta(Q^{\alpha}_{\ \beta})^*\frac{\delta S}{\delta \phi^\alpha} d\mu\,,
%\end{equation}
%where $*$ denotes the formal adjoint of a differential operator (obtained using integration by parts). This is the second Noether theorem, indicating that $\frac{\delta S}{\delta \phi^\alpha}$, the equations of motion of the system, are not all independent. Further insights into the connection between Noether's second theorem and the BV formalism can be found in \cite{FLS03}.

\subsection{Homological interpretation}

Our interest lies in functionals on the solution space $\mathcal{E}_S$ that remain \textit{invariant} under the action $\rho$ of the symmetries. We denote this space as $\mathcal{F}_{S}^{\text{inv}}$ and we use homological algebra to render a description of this space that is more convenient for quantisation using methods discussed in \cite{MichaelChapter}.

\subsubsection{Koszul complex}\label{Kc}

We aim to describe $\mathcal{F}_S$ as the quotient $\mathcal{F}_{S}=\mathcal{F}/\mathcal{F}_0$, where $\mathcal{F}_0$ comprises functionals vanishing on $\mathcal{E}_S$. To characterise such functionals, we use a geometric approach.

The space of solutions $\mathcal{E}_S$ is where the one-form $dS$ vanishes. Consider a vector field $X$. If $X$ possesses suitable locality properties, $\iota_{dS}X\in \mathcal{F}_{0}$.
Let $\Vcal$ denote the space of multilocal polynomial vector fields. Define $\delta_S:\mathcal{V}\rightarrow \mathcal{F}$ by $\delta_S(X)\doteq -\iota_{dS}(X)$. The image of $\delta_S$ is contained in $\mathcal{F}_{0}$. Whether it encompasses all of $\mathcal{F}_{0}$ depends on the system's regularity conditions  \cite{r00075,HT}. See also a recent work \cite{HRV} focusing on the scalar field, which introduces a new class of functionals, suitable for this construction. For a scalar field, the regularity conditions essentially amount to the invertibility of the operator defining the equations of motion.

For a theory with symmetries, one assumes that the system's equations of motion split into independent ones and the ones that can be obtained from them usin g symmetries, rendering the full equation set $dS(\phi)=0$ effectively equivalent to a subset of equations. That subset of equations should then again be described by an invertible local operator. We assume the actions we consider satisfy these regularity conditions.

The kernel of $\delta_S$ comprises vector fields representing symmetries (denoted by $Sym$ below). We establish a chain complex
\[
\begin{array}{c@{\hspace{0,2cm}}c@{\hspace{0,2cm}}c@{\hspace{0,2cm}}c@{\hspace{0,2cm}}c@{\hspace{0,2cm}}c@{\hspace{0,2cm}}c@{\hspace{0,2cm}}c@{\hspace{0,2cm}}c@{\hspace{0,2cm}}c}
	0&\xrightarrow{}&Sym&\hookrightarrow&\mathcal{V}&\xrightarrow{\delta_S}&\mathcal{F}&\rightarrow &0\\
	&&-2&&-1&&0&&&
\end{array}
\]
 and find that the 0th cohomology $H^0=\mathcal{F}/\mathcal{F}_0$, characterizes the space of functionals on the solution space. The negative grading is just a convention.

 From the chain complex above one can obtain a differential graded algebra by taking its graded symmetric powers. We obtain $(\Lambda \Vcal,\delta_S)$, where the differential is extended to higher powers of $\Vcal$ (i.e. multivector fields) by requiring the graded Leibniz rule and setting $\delta_S$ to be zero on $\Fcal$. More explicitly, we write the resulting complex, called \textit{the Koszul complex} as:
 \begin{equation}
\ldots \xrightarrow{\delta_{S}} \underset{-2}{\Lambda^2 \Vcal} \xrightarrow{\delta_{S}} \underset{-1}{\Lambda^1 \Vcal} \xrightarrow{\delta_{S}} \underset{0}{\Fcal } \xrightarrow{\delta_{S}} 0 \ .
\end{equation}
If there are no symmetries, then this complex is a resolution (i.e. only the 0-th cohomology is non-trivial), called \textit{the Koszul resolution}. This is the case, for example, for the real scalar field.

\subsubsection{Chevalley-Eilenberg complex}
Let $\frakg_c$ denote the Lie algebra characterizing infinitesimal local symmetries. By allowing them to act as derivations on functionals with compact support, we can relax the requirement of compact support for the symmetries and instead consider $\frakg$.

We seek the space of symmetry-invariant observables, i.e. those satisfying
\[
\partial_{\rho(\xi)}F=0\,,
\]
for all $\xi\in\frakg$. Algebraically, characterizing the space of invariants under the action of a Lie algebra involves the \textit{Chevalley-Eilenberg complex}.
%%%%%

The Chevalley-Eilenberg complex, denoted by $\CEcal$, is the graded algebra $\Ci_\ml(\overline{\Ecal},\CC)$, representing multilocal functionals on the graded manifold $\Ecal\oplus \frakg[1]\equiv \overline{\Ecal}$ (refer to \cite{Book} for its precise definition) together with a differential $\gamma_{\mathrm{ce}}$. Here, functionals on $\frakg[1]$ correspond to $\Lambda\frakg'$, the exterior algebra over $\frakg'$. Its generators, viewed evaluation functionals, are commonly termed \emph{ghosts}. The grading of $\CEcal$ is referred to as the \emph{pure ghost number} $\#\pg$. We express $\Ci_\ml(\overline{\Ecal},\CC)$ as $\CE\!\doteq\big(\Lambda\frakg'\widehat{\otimes}\F,\gamma_{\mathrm{ce}}\big)$, where $\widehat{\otimes}$ represents the appropriately completed tensor product.

The Chevalley-Eilenberg differential $\gamma_{\mathrm{ce}}$ encodes the action $\rho$ of the gauge algebra $\frakg$ on $\Fcal$. For $F\in\Fcal$, $\gamma_{\mathrm{ce}} F\in \frakg'\widehat{\otimes}\F$ is defined as
\be\label{ChE1}
(\gamma_{\mathrm{ce}} F)(\ph,\xi)\doteq \partial_{\rho(\xi)}F(\ph)\,,
\ee
where $\xi\in\frakg$. In terms of evaluation functionals (i.e., ghosts), $\gamma_{\mathrm{ce}} F=\partial_{\rho(c)} F$. For a form $\omega\in \frakg'$ independent of $\ph$, we set $\gamma_{\mathrm{ce}} \omega(\xi_1,\xi_2)\doteq -\omega([\xi_1,\xi_2])$, resulting in an element of $\Lambda^2\frakg'$. This can be expressed using evaluation functionals as $\gamma_{\mathrm{ce}} c= -\frac{1}{2}[c,c]$. For $F\in\F^\inv$, $\gamma_{\mathrm{ce}} F\equiv 0$, implying $H^0(\CE)$ characterizes gauge invariant functionals.

\subsubsection{BV complex}\label{sec:BV}
In combining gauge invariance and on-shell conditions to characterize the space $\F_S^{\inv}$, we work with the BV complex, denoted $\BV$, representing multilocal compactly supported functionals on the extended configuration space $\overline{\Ecal}$. This space consists of multilocal polyvector fields on $\overline{\Ecal}$, encompassing field multiplets $\ph^\alpha$ and corresponding antifields $\ph_\alpha^\ddagger$. 
 We distinguish between right $\frac{\delta_r }{\delta \ph^\alpha}$ and left derivatives $\frac{\delta_l }{\delta \ph^\alpha}$. We conventionally identify antifields with right derivatives.

The algebra $\BVcal$ has two gradings: the ghost number $\#\gh$ (the main grading) and the antifield number $\#\af$ (an additional grading used later). Functionals of physical fields have both numbers equal to 0. Functionals of ghosts have $\#\af=0$ and $\#\gh=\#\pg$ (the \emph{pure ghost} grading, where a ghost $c$ has $\#\pg=1$). All vector fields have a non-zero antifield number given by $\#\af(\ph_\alpha^\ddagger)=1+\#\pg(\ph^\alpha)$, and $\#\gh=-\#\af$.

$\BVcal$, viewed as the space of graded multivector fields, is equipped with a graded generalization of the Schouten bracket, called \textit{the antibracket}. The right derivation $\delta_S$ is not inner with respect to $\{.,.\}$, but locally it can be written as:
\[\delta_SX=\{X,L(f)\}, \quad f\equiv 1 \ \textrm{on} \  \supp X, \ X\in \V.\]
We denote this as $\delta_SX=\{X,S\}$. Similarly, one can find an action $\theta$ such that $\gamma_{\mathrm{ce}} X=\{X,\theta\}$, and we define the \textit{classical BV differential} as
\[
s=\{.,S+\theta\}\equiv\{.,S^{\ex}\}\,,
\]
where $S^{\ex}$ is the \textit{extended action}. The BV differential $s$ must be nilpotent, i.e., $s^2=0$, which leads to the \textit{classical master equation} ({\cme}):
\be\label{CME}
\{L^{\ex}(f),L^{\ex}(f)\}\sim 0,
\ee
with respect to the equivalence relation \eqref{equ}.

The differential $s$ increases the ghost number by one (i.e., it has order 1 in $\#\gh$). It can be expanded with respect to the antifield number as
\[
s=\tilde{\delta}+\tilde{\gamma},
\]
where $\tilde{\delta}$ has order -1 in $\#\af$ and extends $\delta_S$, while $\tilde{\gamma}$ has order 0 and extends $\gamma_{\mathrm{ce}}$.
This results in the following bicomplex structure:
\[
\begin{CD}
\ldots@>\tilde{\delta}>>\big(\La^2\V\oplus\fG\big) @>\tilde{\delta}>>\V@>\tilde{\delta}>>\F@>\tilde{\delta}>>0\\ 
@.     @VV{\tilde{\gamma}}V@VV{\tilde{\gamma}}V@VV{\tilde{\gamma}}V@.\\
\ldots@>\tilde{\delta}>>{\big(\La^2\V\oplus\fG\big)\widehat{\otimes}\frakgo'}@>\tilde{\delta}>>{\Vcal\widehat{\otimes}\frakgo'}@>\tilde{\delta}>>{\Fcal\widehat{\otimes}\frakgo'}@>\tilde{\delta}>>0
\\ 
@.     @VV{\tilde{\gamma}}V@VV{\tilde{\gamma}}V@VV{\tilde{\gamma}}V@.\\
@.\dots @.\dots@.\dots@.\\ 
\end{CD}
\]
Here $\fG$ is the space of $\frakg_c$-valued multilocal functionals on $\Ecal$ and it characterises the space of local gauge symmetries, i.e. it is added in degree $-2$ to compensate for the kernel of $\tilde{\delta}$ in degree $-1$. The restriction to compact support matches the fact that $\Vcal$ is the space of \textit{compactly supported} vector fields, so in all this discussion we treat only \textit{local} gauge symmetries. This makes sense, since these are exactly the ``problematic'' symmetries that obstruct the equations of motion from being globally hyperbolic.

We assume that $\tilde{\delta}$ gives us a resolution. If not, one needs to add higher-order terms to the differential, but we will not discuss this here, as this is not needed for Yang-Mills theory.
Crucially, we have 
\be\label{eq:physical}
H^0(\BV,s)=\F_S^{\inv},
\ee
which is the reason for working with $\BV$ in the first place, as it contains the same information as $\F_S^{\inv}$ but has a simpler algebraic structure (quotients and spaces of orbits are resolved).

In the next step, we introduce the gauge fixing. For this purpose we extend the BV complex with antighosts $\bar{c}$ (in degree -1) and Nakanishi-Lautrup fields $b$ (in degree 0). These form a trivial pair, i.e.:
	\[
	s\bar{c}^I=ib^I,\quad s b^I=0.
	\]
	The new extended configuration space is written explicitly as
	\[
	\overline{\Ecal}=\Ecal\oplus\frakg[1]\oplus\frakg[0]\oplus\frakg[-1].
	\]
	Since the new generators were introduced as a trivial pair, the cohomology of the resulting complex is the same as the original one, so \eqref{eq:physical} remains true also after this modification. We now use %using 
 an automorphism $\alpha_\Psi$, defined on generators as 
\[
\alpha_\Psi(\ph^\ddagger_\beta(x))\doteq \ph^\ddagger_\beta(x)+\frac{\delta \Psi(f)}{\delta \ph^\beta(x)}, \quad \alpha_\Psi(\ph^I(x))=\ph^I(x),
\] 
where $f(x)=1$ and $\Psi_M(f)$ is a fixed generalized Lagrangian of ghost number -1 which does not contain antifields, called \textit{gauge fixing fermion}, and we choose it in such a way that the antifield free part of the transformed action gives rise to hyperbolic equations (see \cite{FR} for details). This redefinition is a canonical transformation with respect to the antibracket and we still have that $\Fcal^{\inv}_S=H^0(\BV,s)$, where $s$ now uses the antibracket with the appropriately redefined action. The difference is that now we can expand $s$ with respect to a different grading, namely the \textit{total antifield number} $\#\ta$, which is 1 for all the antifield generators and zero for fields. We need that different expansion, since now our equations of motion are normally hyperbolic, so we don't need to include symmetries in degree $-2$ of the horizontal complex. With this different book-keeping, all the antifields end up in degree $-1$. We write
\[
s=\delta+\gamma,
\]
where we again use the notation $\delta$ and $\gamma$ for the two terms in the expansion.

Crucially, $\delta$ now describes the \emph{gauge-fixed} equations of motion, and is again a resolution, so
\[
\Fcal^{\inv}_S=H^0(\BV,s)=H^0(H_0(\BV,\delta),\gamma).
\]
Taking $H_0(\BV,\delta)$ is interpreted as \emph{going on-shell}. This cohomological interpretation of gauge-fixing has been discussed in \cite{BHHS}.

%%%%%%%%%%%%%%%%%%%%%%%%%%%
\begin{exa}
Consider the Yang-Mills theory. In the view of gauge-fixing, we introduce Nakanishi-Lautrtup fields and antighosts, so we now have the following types of fields:
\begin{itemize}
    \item the vector potential $A\in\Omega^1(M,\mathfrak{k})$, with degree 0.
    \item the ghost $c\in\cinfty(X,\mathfrak{k})$, with degree 1, 
    \item the antighost $\bar{c}\in\cinfty(X,\mathfrak{k})$, with degree -1, and
    \item the Nakanishi-Lautrup field $b \in\cinfty(X,\mathfrak{k})$, with degree~$0$.
\end{itemize}
We also introduce antifields for all the objects above and if a field $\ph$ is of degree $m$, 
then its antifield $\ph^\ddagger$ has degree~$-m-1$, e.g. the vector potential $A$ has antifield $A^\ddagger$ of degree~$-1$.

To specify the generalized Lagrangian~$L_{YM}$,
we need to specify how it depends on test functions. For classical theory we only need the classical master equation to hold in the weak sense \eqref{CME}, but in the view of quantisation we need something stronger.

We fix a pair of test functions $f=(f_A,f_c)$ and in the expression for the Yang-Mills Lagrangian density we multiply the vector potential $A$ by $f_A$ and the ghost $c$ by $f_c$. We also impose the condition that 
\[
\label{eq: supp for YM}
f_A|_{\supp(f_c)} \equiv 1
\]
so that the gauge transformation associated to $f_c c$ has compact support (generically) within the support of~$f_A A$.
%(In the remark below, we expand upon the reason for this condition.)

%We use the following notation for vector fields (see \cite{FR3}): a vector field $V$ that acts on functions as $\partial_V F(\ph)= \int_X V(\ph) \delta F|_{\phi}\equiv  \int_X V(\ph)(x) \frac{\delta F}{\delta \phi(x)}$ is denoted by $\int_X V(\ph)(x) \frac{\delta}{\delta \phi(x)}$. This is to emphasize the fact that $V(\ph)(x)$, for $\phi\in\mathcal{E}$, can be thought of as a set of coordinates and the differential operators $\frac{\delta}{\delta \phi(x)}$, called antifields, act as basis for $T_\ph \mathcal{E}$.

The generalized Lagrangian including the non-minimal sector is
\begin{multline*}
L_{YM}(f)
=-\frac{1}{2}\int_X \tr\big(F[f_A\, A]\wedge *F[f_A\, A]\big)
+\int_X \big(d(f_c\,c)+\frac{1}{2}[A,f_c\,c]\big)^I_\mu(x)\,{A^{\ddagger}}_I^\mu(x)\\
+\frac{1}{2}\int_X  [c,f_c\,c]^I(x)\,c_I^\ddagger(x)
-i\int_X  f_c\,b_I(x)\, (\bar{c}^{\ddagger})^I(x)\,,
\end{multline*}
when written in local coordinates and a fixed basis for~$\mathfrak{k}$. Direct computation verifies that this Lagrangian satisfies the strict version of CME, i.e.
\[\{L_{YM}(f),L_{YM}(f)\}=0\,.
\]
\end{exa}
\begin{exa}
In general, we use the following prescription for constructing a Lagrangian that satisfies CME in the strict sense. Consider a theory with fields $\ph^{\alpha}$, $\alpha=1,\dots N$, ghosts $c$ and the non-minimal sector consisting of $b,\overline{c}$.
\begin{itemize}
    \item Choose $f_c,f_b\in \Dcal$ and $f_{\bar{c}},f_{\alpha}\in\Dcal$, $\alpha=1,\dots, N$, so that  $f_{\alpha}\equiv 1$  on the support of $f_c$ for all $\alpha=1,\dots N$, and $f_{\bar{c}}\equiv 1$ on the support of $b$.
    \item Replace all the fields $\ph^\alpha$ in the Lagrangian density with $f_{\alpha}\ph^{\alpha}$.
    \item Replace the antifields $\ph^\ddagger\equiv \frac{\delta}{\delta \ph^{\alpha}}$ with $\frac{\delta}{\delta (f_{\alpha} \ph^{\alpha})}$, the latter understood as the operation of differentiation by $\ph^\alpha$ followed by division by $f_{\alpha}$. The same for antifields of ghosts, antighosts and $b$ fields. 
    \item Check if the resulting expression is well-defined. After taking the support properties into account, one should be able to write the resulting smeared Lagrangian in a form where no divisions by test functions are present.
    \item Often one can choose some of the test functions to be the same, so for example for Yang-Mills theory we only needed two.
\end{itemize}
This prescription has been applied to the Einstein-Hilbert action in \cite{BFRej13}. The advantage of this prescription is that one can use the algebraic relations that would hold for the Lagrangian when restricted to compactly-supported fields to show that the CME holds in the strict sense.
\end{exa}

%%%%%%%%%%%%%%%%%%%%%%%%%%%%

\begin{exa}
	To implement a Lorenz-like gauge in Yang-Mills theory, 
  % we need to further extend the BV complex with antighosts $\bar{C}$ (in degree -1) and Nakanishi-Lautrup fields $B$ (in degree 0). These form a trivial pair, i.e.:
	% \[
	% s\bar{C}^I=iB^I,\quad s B^I=0.
	% \]
	% The new extended configuration space is written explicitly as
	% \[
	% \overline{\Ecal}=\Ecal\oplus\frakg[1]\oplus\frakg[0]\oplus\frakg[-1].
	% \]
	% Since the new generators were introduced as a trivial pair, the cohomology of the resulting complex is the same as the original one, so \eqref{eq:physical} remains true also after this modification. 
 we choose the gauge-fixing fermion as
\[
\Psi(f)=i\int\limits_X f_A\,\bar{c}_I\left(\frac{1}{2}f_c\, b^I+*^{-1}d*\!(f_A\, A^I)\right)\dvol\,,
\]
where $f=(f_A,f_c)$.
The gauge-fixed action is 
\[
L(f)=L_{YM}(f)+\{L_{YM}(f),\Psi(f)\}\,.
\]
This action also satisfies the strict version of CME.
\end{exa}

\subsubsection{Linearized theory}
We decompose the extended action into two components: $S_0$, quadratic in both fields and antifields, and the interaction term $V$. Specifically, $S_0$ is expressed as:
\[
S_0=S_{00}+\theta_0\,,
\]
where $S_{00}$ denotes the part with $\#\ta=0$, and $\theta_0$ pertains to $\#\ta=1$. Likewise, we represent $V$ as $V=V_{0}+\theta$, and observe that $S=S_{00}+V_{0}$ represents the entirety of the action independent of antifields.

We introduce the linearized BRST differential
\[
\gamma_0 F\doteq \{F,\theta_0\}\, .
\]
The total linearized BV differential $s_0$ is
\[
s_0=\delta_0+\gamma_0\,,
\]
where $\delta_0(\ph^\ddagger_\alpha)=-\frac{\delta S_{00}}{\delta\ph^\alpha}$. Consequently, the homology of $\delta_0$ characterizes the space of solutions to the linearized equations of motion. We denote:
$$\frac{\delta_l S_{00}}{\delta\ph^\al(x)}(\ph)\equiv P_{\alpha\beta}(x)(\ph^\beta(x))\,,$$
where each component $P_{\alpha\beta}$ represents a differential operator. For conciseness, we often express the equations of motion using index-free notation: $P\ph=0$.

Similarly, we define:
$$\frac{\delta_r\delta_l \theta_0}{\delta\ph^\sigma(y)\delta\ph_\al^\dgr(x)}\equiv K^{\al}_{\phantom{\al}\sigma}(x)\delta(y-x)\,,$$
where each $K^{\alpha}_{\ \sigma}$ denotes a differential operator.

The cohomology of $s_0$ is given by $H^0(\BV,s_0)=H^0(H_0(\BV,\delta_0),\gamma_0)$,
since $(\BV,\delta_0)$ is a resolution. Taking $H_0(\BV,\delta_0)$ is understood as \emph{going on shell}.

Assume that the gauge fixing was done in such a way that $P$ is Green hyperbolic (for gauge theories and gravity this was shown in \cite{FR}), meaning that there exist unique retarded and advanced Green functions $\Delta^{\rm A/R}$, i.e. Green functions for the equations of motion operator $P$ such that
\[
\supp(\Delta^{\rm R}(f))\subset  J^+(\supp(f))\,,\qquad
\supp(\Delta^{\rm A}(f))\subset J^-(\supp(f))\,.
\]
We define the Pauli-Jordan function by
\[
\Delta=\Delta^{\rm R}-\Delta^{\rm A}\,.
\]

The classical linearized theory is constructed by introducing the Peierls bracket given by:
\begin{equation}\label{eq:Peierls}
\Pei{F}{G} = \sum_{\al,\beta} \skal{\frac{\delta^r F}{\delta\ph^\al}}{{\De}^{\al\beta}\frac{\delta^l G}{\delta\ph^\beta}},
\end{equation}
where $F, G \in\BV$. Unfortunately, $\BV$ is not closed under this bracket and one needs to extend it to a larger space. A good candidate is the space $\BV_\mc$ of \textit{microcausal} polynomials on $T^*[-1]\overline{\Ecal}$, i.e. polynomial functionals that are smooth, compactly supported and their derivatives (with respect to both $\ph$ and $\ph^\dgr$) satisfy the WF set condition:
\be\label{mlsc}
\WF(F^{(n)}(\ph,\ph^\dgr))\subset \Xi_n,\quad\forall n\in\NN,\ \forall\ph\in\overline{\E}(M)\,,
\ee
where $\Xi_n$ is an open cone defined as 
\be\label{cone}
\Xi_n\doteq T^*M^n\setminus\{(x_1,\dots,x_n;k_1,\dots,k_n)| (k_1,\dots,k_n)\in (\overline{V}_+^n \cup \overline{V}_-^n)_{(x_1,\dots,x_n)}\}\, .
\ee
Compare with \cite{MichaelChapter}. For the discussion of non-polynomial functionals, one needs to replace microcausal functionals with a smaller space that is closed under the Peierls bracket, e.g. the \emph{equicausal functionals} proposed in \cite{HRV}.

\subsection{Quantization}
\subsubsection{Free theory}\label{free:theory} 
The quantized algebra of free fields is constructed by means of \textit{deformation quantization} of the classical algebra $(\BV_{\mc},\Pei{.}{.})$. The procedure is exactly as in \cite{MichaelChapter}, but for a larger field multiplet and with some grading involved. The $\star$-product is
\begin{equation*}%\label{star product}
F\star G\doteq m\circ \exp({i\hbar D_W})(F\otimes G),
\end{equation*}
where $m$ is the multiplication operator, i.e. $m(F\otimes G)(\ph)=F(\ph)G(\ph)$, and  $D_W$ is the functional differential operator defined by
\begin{equation*}%\label{star product2}
D_W\doteq \frac{1}{2} \sum_{\al, \beta} \left<{W}^{\al\beta},\frac{\delta^l}{\delta\ph^\al} \otimes \frac{\delta^r}{\delta\ph^\beta}\right>\,.
\end{equation*}
with $W$, the \textit{2-point function of a Hadamard state}. $W$  satisfies the appropriate wavefront set condition \cite{Rad} and we have \textit{$W=\frac{i}{2}\Delta+H$}, where $H$ is a symmetric bisolution for $P$. 
In order to achieve that $\gamma_0$ is a right derivation of the $\star$-product,
\[
\gamma_0(X\star Y)=(-1)^{\#\gh(Y)}\gamma_0 X\star Y+X\star \gamma_0Y\ ,
\]
%In addition to these standard properties, 
we also require the\textit{ consistency condition }\cite{H} on the symmetric part:
\be\label{const:cond2}
\sum_\sigma((-1)^{|\ph^\al|}K^{\al}_{\ \sigma}(x')H(x',x)^{\sigma\gamma}+K^{\gamma}_{\ \sigma}(x)H(x',x)^{\al\sigma})=0\, .
\ee
% Under this condition, $\gamma_0$ is a right \todo{what is a right derivation?} derivation with respect to the star product. 
Since $W$ is a solution for the linearized equations of motion operator $P$, $\delta_0$ is also a right derivation with respect to $\star$. We can therefore conclude that the same holds for $s_0$,
\[
s_0(X\star Y)=(-1)^{\#\gh (Y)}s_0X\star Y+X\star s_0Y\,.
\]
\subsubsection{Interacting theory}
In our exposition, we commence with an analysis of \textit{regular polynomial functionals} in $\BVcal$, denoted by $\BV_{\reg}$. The \textit{time-ordering operator $\TT$} is defined as follows:
\[
\TT F(\ph)\doteq e^{\frac{\hbar}{2}\Dcal_{\Delta^{\rm F}}}\ ,
\]
where $\Delta^{\rm F}=\frac{i}{2}(\Delta^{\rm A}+\Delta^{\rm R})+H$ and for an integral kernel $M$, we introduce
$$\Dcal_M\doteq \sum_{\alpha,\beta}\left<{M}^{\alpha\beta}, \frac{\delta^l}{\delta\ph^\alpha}\frac{\delta^r}{\ph^\beta}\right>\,.$$
 
Formally, $\TT$ corresponds to the operator of convolution with the ``oscillating Gaussian measure with covariance $i\hbar\Delta^{\rm F}$''
\[
\TT F(\ph)\overset{\text{formal}}{=} \int F(\ph-\phi)\, d\mu_{i\hbar\Delta_F}(\phi) \ . 
\]
We define the \textit{time-ordered product} $\T$ on $\BV_\reg[[\hbar]]$ by:
\[
F\T G\doteq \Tcal(\Tcal^{\minus}F\cdot\Tcal^{\minus}G)
\]

The time-ordered product $\T$ is the time-ordered version of the star product $\star$, namely $F\T G=F\star G$ if the support of $F$ is not earlier than the support of $G$, and $F\T G=G\star F$ if the support of $G$  is not earlier than that of $F$.

We represent interactions through functionals $V$, initially assuming $V\in\BV_\reg$. The quantum observable associated with $V$ is denoted as $\TT V$. Analogously to normal ordering, we denote $\TT V\equiv \no{V}$.

The \textit{formal S-matrix}, denoted as $\Scal(\lambda \no{V})\in\BV_{\reg}((\hbar))[[\lambda]]$, is defined by:
\[
\Scal(\lambda V)\doteq e_{\sst{\TT}}^{i\lambda\no{V}/\hbar}=\TT(e^{i\lambda V/\hbar})\,.
\]

\textit{Interacting fields} belong to $\BV_{\reg}[[\hbar,\lambda]]$ and are given by
\[
R_{\lambda V}(F)\!\doteq\! (e_{\sst{\TT}}^{i\lambda \no{V}/\hbar})^{\star \minus}\star (e_{\sst{\TT}}^{i\lambda \no{V}/\hbar}\T \no{F})=-i\hbar\frac{d}{d\mu}\Scal(\lambda V)^{-1}\Scal(\lambda V+\mu F) \bigg|_{\mu=0}
\]

For $\lambda=0$, $R_{0}(F)=\no{F}$. The \textit{interacting star product} is defined as:
\[
F\star_{\text{int}} G\doteq R_V^{\minus}\left(R_V(F)\star R_V(G)\right)\,.
\]

The problem that one faces is that interesting interactions and observables are local, but not regular. Because of singularities of $\De^{\rm F}$, the time-ordered product $\T$ is not well defined on local, non-linear functionals, but the \textit{physical interactions are usually local}!

The  \textit{renormalization problem} is then to extend $\Scal$ to local arguments by extending time-ordered products:
\[
\Scal(V)=\sum\limits_{n=0}^\infty \frac{1}{n!} \TT_{n}(V,...,V)\,.
\]
This is addressed in \cite{MichaelChapter}.

\subsubsection{{\qme} and the quantum BV operator}

In the framework of \cite{FR3}, a fundamental condition is the invariance of the S-matrix under the free classical BV operator:
\begin{equation}\label{invmatrix}
	s_0\left(e_{\sst{\TT}}^{i \no{V}/\hbar}\right)=0\,,
\end{equation}
where $\TT$ denotes the time-ordering operator. A key identity satisfied by $\TT$ is:
\begin{equation}\label{qbv1}
	\delta_{0}(\TT F)=\TT(\delta_0 F-i\hbar\Lap F)\,,
\end{equation}
with $\Lap$ representing the BV Laplacian. Additionally, from the consistency conditions \eqref{const:cond2}, we derive:
\begin{equation}\label{qbv2}
	\TT\circ \gamma_0=\gamma_0\circ \TT\ .
\end{equation}
Combining these, setting $\Lap L_0(f)=0$ and employing the classical master equation yields:
\[
s_0\left(e_{\sst{\TT}}^{i \no{V}/\hbar}\right)=\frac{i}{\hbar}\,e_{\sst{\TT}}^{i \no{V}/\hbar}\T \TT\left(\frac{1}{2}\{L(f),L(f)\}-i\hbar\Lap(L(f))\right)\,,
\]
where $L=L_0+V$. This implies that condition \eqref{invmatrix} is equivalent to the quantum master equation ({\qme}):
 \be\label{QME}
\frac{1}{2}\{L(f),L(f)\}-i\hbar\Lap(L(f))=0\,.
\ee
Note that fulfilling this equation might require a particular choice of how the Lagrangian density is smeared with the test function $f$ or a family of test functions, as discussed at the end of Section~\ref{sec:BV}. This is in contrast with the {\cme}, which was only required to hold in the weaker form \eqref{CME}.

In the free theory, the quantum BV operator is just:
\begin{equation}\label{quantumBV}
	\hat{s}_0\doteq \TT^{-1}\circ s_0 \circ \TT\,,
\end{equation}
thus, from \eqref{qbv1} and \eqref{qbv2}:
\[
\hat{s}_0=s_0-i\hbar \Lap\,.
\]
In the interacting theory, $\hat{s}$ is defined on regular functionals by:
\[
\hat{s}=R_V^{-1} \circ s_0\circ R_V\,,
\]
and characterizes quantum gauge invariant observables. Assuming \qme, we have:
\[
\begin{aligned}
	\hat{s}F=& \{F,S_0+V\}-i\hbar \Lap(F)=s-i\hbar \Lap(F) \,.
\end{aligned}
\]
To extend {\qme} and $\hat{s}$ to local observables, I replace now $\T$ with the renormalized time-ordered product, as proposed in \cite{FR3} and use the  \textit{anomalous Master Ward Identity} (AMWI) \cite{BreDue,H} to obtain the \emph{renormalized} {\qme}.
\[
\tfrac{1}{2}\{L(f),L(f)\}-i\hbar A_{L(f)}=0\,,
\]
where $A_{L(f)}$ is the \emph{anomaly term}, which is local and depends locally on $L(f)$. Replacing $V$ with $V+\lambda F$ in the AMWI and differentiating with respect to $\lambda$ leads to the following result for  the renormalized BV operator:
\[
			\hat{s}F=\{F,S_0+V\}-i\hbar\Lap_V(F)\,,
\]
where $\Lap_V F\doteq \frac{d}{d\lambda} \Lap_{V+\lambda F}\big|_{\lambda=0}$.

 Hence, by using the renormalized time ordered product, we obtained in place of $\Lap$ (which is ill-defined on local vector fields), the interaction-dependent operator $\Lap_V$. It is of order $\mathcal{O}(\hbar)$ and it is local.
 
 There are some subtleties here that we want to point out. First is the interplay between compact support of functionals and non-compact support of fields. Since the functionals are always compactly supported, they are not sensitive to what happens outside a given local region. Also the symmetries that are being quotiented out in our analysis are only the compactly supported symmetries (i.e. compactly supported vector fields). This is in contrast to the approach of \cite{MollerMaps}, where the authors also include global symmetries and show the non-existence of M{\o}ller maps intertwining the free and the interacting theory. Here we construct such maps explicitly with the caveat that the resulting interacting quantum BV operator might be non-local. Its locality requires us to assume the quantum master equation in the form \eqref{QME}.

\section{Non-perturbative formulation}
The algebraic relations valid in pAQFT can actually be used to define directly C*-algebras which then constitute an algebraic quantum field theory in the original framework of Haag and Kastler. For an interacting scalar field this was analysed  
in a recent work by Buchholz and Fredenhagen \cite{BF19}. 
% demonstrated the feasibility of formulating an interacting quantum theory for the scalar field using $C^*$-algebras. 
This formulation involves defining local S-matrices $\Scal$ as unitary operators indexed by local functionals, and constructing a $C^*$-algebra generated by those unitaries. Subsequently, relations are imposed to ensure that the S-matrices adhere to rules motivated by pAQFT \cite{BuchholzFredenhagenEnzyclopedia}.

Let $F_1,F_2$ be local functionals and let $F_1\prec F_2$ denote the relation: $\supp F_1$ is not to the future of $\supp F_2$ (i.e. $\supp F_1$ does not intersect $J^+(\supp F_2)$). Local S-matrices are required to satisfy:
	\begin{enumerate}[{\bf S1}]
		\item {\bf Identity preserving}:\label{S:start} $\Scal(0)=\1$.
		\item {\bf Locality}:\label{S:loc} $\Scal$ satisfies the causal factorization property, i.e. 
		$F_1\prec F_2$ implies that
		\[
		\Scal(F_1+F+F_2)=	\Scal(F_1+F)\Scal(F)^{-1}	\Scal(F+F_2)\,,
		\]
		where $F_1, F, F_2\in \F_\loc$.
	\end{enumerate}

Following \cite{BF19}, let $\fA$ denote the group algebra over $\CC$ generated by elements $\Scal(F)$, where $F\in \Fcal_{\loc}$, modulo the relations {\bf S\ref{S:loc}} and {\bf S\ref{S:start}}. Furthermore, given a fixed $L\in \euL$ (interpreted as the Lagrangian of the theory), one defines $\fA_{L}$ by additionally quotienting by the relation proposed by \cite{BF19} to encode the dynamics:
\begin{enumerate}[{\bf S1}]
 \addtocounter{enumi}{2}
\item \[%\label{eq:SD}%\tag{\bf S3}
\Scal(F)
% \Scal(\delta L(\ph))
=\Scal(F^{\ph}+\delta L(\ph))\,,
% =\Scal(\delta L(\ph))\Scal(F),
\]
\end{enumerate}
where $F^{\ph}(\psi)\doteq F(\ph+\psi)$, $\ph,\psi\in \Ecal$, and $\delta L$ is defined as in Definition~\ref{df:delta:L}. Physically, condition {\bf S3} is interpreted as the \textit{Schwinger-Dyson equation} on the level of local S-matrices.

Further developments involve the generalisation to Fermions \cite{BDFR22}, the characterisation of the renormalisation group and the formulation of the unitary anomalous master Ward Identity (UAMWI) \cite{BDFR23}.

The relations above determine the algebraic structure for the case of a quadratic Lagrangian $L$ and linear functionals $F$ and yield an algebra isomorphic to the Weyl algebra which is known to be simple. This is not true for the full algebra. The freedom in introducing additional relations can to some extent be incorporated in a group $R$ acting on local functionals which leaves linear functionals invariant. This group may be compared to the renormalisation group of Petermann and Stückelberg which according to Stora's Main Theorem of Renormalisation characterizes the freedom in the choice of renormalisation conditions in perturbative quantum field theory \cite{PS82}. In particular the group $R$ appears in the description of anomalies in the action of symmetries $g$ on the classical configuration space. The anomalies modify the induced action of the symmetry on local functionals by a cocycle with values in the  group $R$. Given such a cocycle $g\mapsto\zeta_g$, the UAMWI is imposed as an additional relation,
\be
S(\zeta_g F)=S(g_*F+\delta_gL)
\ee
where $\delta_gL$ denotes the change of the action induced by the symmetry $g$.
This relation allows one to prove the time-slice axiom and to describe the action of global symmetries of the Lagrangian in terms of a version of the Noether Theorem.
Moreover, the flow of Lagrangians under scale transformations ({\it running coupling constants}) can be obtained. This motivates to name $R$ the {\it renormalisation group} in this nonperturbative framework.

The derived structure is in agreement with the structure found in perturbation theory, with two modifications: in perturbation theory the cocycle characterizing the anomalies is, up to equivalence by finite renormalisations, uniquely determined as a consequence of the Main Theorem of Renormalisation, but in the abstract algebraic framework, it is an additional datum which can be chosen freely. The other difference concerns the notion of causality. In perturbation theory the causal structure of the free theory is not modified by the interaction, but in the nonperturbative case changes of the causal structure are possible. Namely, in the causal factorization condition above the causal relation between the supports of the functionals $F_1,F_2$ refers to the causal structure induced by the interaction $F$.

Compared to other attempts towards a nonperturbative construction of interacting quantum field theories, the remaining open problem is whether the constructed algebra has states with a suitable physical interpretation as {\it e.g.}\ vacuum states, particle states etc.. For the subalgebra generated by local functionals of 2nd order it could be shown that an extension of the vacuum representation of the free field is possible \cite{BF-Kin}. 

% Crucially, the formulation of the UAMWI requires introducing a renormalisation group, which in this setting is also defined abstractly, as a certain group acting on the space of local functionals.\todo{Quantum Noether theorem, non-perturbative renorm goup}

\bibliographystyle{amsalpha}
\bibliography{References}

\end{document}